\documentclass[10pt,conference]{IEEEtran}
\usepackage{amssymb}
\usepackage{graphicx}
\usepackage{url}
\newtheorem{theorem}{Theorem}
\newtheorem{rom}{Theorem}

\newtheorem{definition}{Definition}

\newtheorem{lemma}{Lemma}

\newtheorem{remark}{Remark}

\begin{document}

\title{On the Construction for Quantum Code $((n,K,d))_{p} $
\\via Logic Function over ${\rm {\mathbb F}}_{p}$}

\author{
\authorblockN{Shuqin Zhong, Zhi Ma, Yajie Xu and Xin L$\ddot{u}$
 }
\authorblockA{ Zhengzhou Information Science and Technology
Institute  \\
Zhengzhou, 450002, China\\
Email: lavenderzhong@live.cn} 
}
 \maketitle

\begin{abstract}
This paper studies the construction for quantum codes with
parameters $((n,K,d))_{p}$ by use of an \textit{n}-variable logic
function with APC distance $d'\ge 2$ over ${\rm {\mathbb F}}_{p} $,
where $d$ is related to $d'$.  We obtain $d\le d'$ and the maximal
$K$ for all $d=d'-k$, $0\le k\le d'-2$. We also  discuss the basic
states and the equivalent conditions of saturating quantum Singleton
bound.
\end{abstract}

\section{Introduction}
Quantum error correcting code \cite{proceeding1},
\cite{proceeding2}, \cite{proceeding3}, \cite{proceeding4} has
become an indispensable element in many quantum information tasks
such as the fault-tolerant quantum computation \cite{proceeding5}
the quantum key distribution \cite{proceeding6} and the entanglement
purification \cite{proceeding7}, \cite{proceeding8},  to fight the
noises.

Early in 1998, Calderbank \cite{proceeding9} presented systematic
mathematical methods to construct binary quantum codes (stabilizer
codes) from classical error correcting codes over ${\rm {\mathbb
F}}_{2} $ or ${\rm {\mathbb F}}_{4} $. A series of good binary
quantum codes were constructed by using classical codes (BCH codes,
Reed-Muller codes, AG codes, etc.). Schlingemann and Werner
\cite{proceeding10}  proposed a new way to construct quantum
stabilizer codes by finding certain graphs (or matrices) with
special properties. Using this method they constructed several new
non-binary quantum codes. In particular, they gave a new proof on
the existence of quantum code $ [[5,1,3]]_p $ for all odd primes $p$
(the first proof was given by Rain \cite{proceeding11}). It seems
that this method can be used to obtain many quantum codes saturating
quantum Singleton bound (For any code $ [[n,k,d]]_p $ , the quantum
Singleton bound says that $n\ge k+2d-2$, see \cite{proceeding3} for
$ p= 2$ and \cite{proceeding11} for \textit{$p\ge 3$}). We call this
kind of quantum codes quantum MDS codes.  At the same time, Feng
Keqin \cite{proceeding12} showed there existed quantum codes $
[[6,2,3]]_p $ and $ [[7,3,3]]_p $ for any prime number $p$. Liu
Tailin \cite{proceeding13} proved the existence of quantum codes $
[[8,2,4]]_p $ and $ [[n,n-2,2]]_p $ for all odd prime numbers $p$.

In the correspondence, researchers made use of Boolean functions and
projection operators \cite{proceeding14} to find quantum error
correcting codes. In Ref \cite{proceeding15}, the author constructed
quantum code with parameters $ [[n,0,d]]_p $, where $d$ is the APC
distance of a Boolean function. Xu \cite{proceeding16} generalized
the definition of APC distance for  Boolean functions to logic
functions over ${\rm {\mathbb F}}_{p} $, then constructed quantum
code $((n,K,d))_{p} $, where $d$ is related to APC distance of an
\textit{n}-variable function over ${\rm {\mathbb F}}_{p} $. Before
talking further more about the ideas and results of this paper, we
need to introduce the logic construction of Ref \cite{proceeding16}
which will be used in this paper.

For $d'\ge 2$, let $f(x)$ be a function with $n$ variables and APC
distance $d'$ over ${\rm {\mathbb F}}_{p} $. $\beta _{i}
=\left(\beta _{i1} ,\cdots ,\beta _{in} \right)\in \rm {\mathbb
F}_{p}^{n} $ for all $1\le i\le K$.

\begin{lemma}\cite{proceeding16}
The space spanned by $\{ |\psi _{i} \rangle =p^{-\frac{n}{2} } \sum
_{x\in {\rm {\mathbb F}}_{p}^{n} } \zeta ^{f(x)+\beta _{i} x}
|x\rangle |1\le i\le K\} $ is a quantum code with parameters
$((n,K,d))_{p} $ satisfying:

\[d=min\{ W_{s} (u,v)|\exists 1\le i\le j\le K,W_{s} (u,v-\beta _{i} +\beta _{j} )\ge d'\}, \]
\noindent where $ \zeta $ is a primitive element in $ \mathbb{F}_p
$.
\end{lemma}

This result was proved by Xu in \cite{proceeding16}. Following the
work of Xu, we discussed the parameters and basic states of the
constructed quantum code. The main results proved in this paper are:

\begin{theorem} Quantum code $((n,K,d))_{p} $ spanned by \[\{ |\psi _{i} \rangle =p^{-\frac{n}{2} } \sum
_{x\in {\rm {\mathbb F}}_{p}^{n} } \zeta ^{f(x)+\beta _{i} x}
|x\rangle |1\le i\le K\} \]  is with following properties:
\begin{enumerate}
\item  $d\le d'$,

\item  $\beta _{1} =\cdots =\beta _{K} =0\; $ for $d=d'$,

\item  $W_{H} \left(\beta _{i} ,\beta _{j} \right)\le k$ for all $d'=d-k$ if $0<k\le d'-2$.
\end{enumerate}
 \end{theorem}

\begin{theorem}
If quantum code $((n,K,d))_{p} $ is spanned by \[\{ |\psi _{i}
\rangle =p^{-\frac{n}{2} } \sum _{x\in {\rm {\mathbb F}}_{p}^{n} }
\zeta ^{f(x)+\beta _{i} x} |x\rangle |1\le i\le K\}. \] Then,

\[K=\left\{\begin{array}{l} {1\; ,\; \; \; \; \; \; d=d'} \\ {\le p,\; \; \; d=d'-1} \\ {\le \max p^{k-2} (1+n(p-1),p^{2} )\; \; ,\; d=d'-k} \end{array}\right. ,\]
where $2\le k\le d'-2$.
\end{theorem}

We state the logic description of quantum codes in Section II and
the proof of our main results in Section III . Section IV is largely
devoted to the basic states and equivalent conditions of
constructing quantum codes saturating quantum Singleton Bound.
Conclusions are drawn in Section V.

\section{A Logic Description of Quantum Codes}
The logic description of quantum codes given by \cite{proceeding16}
can be stated in following element way.

Let $f(x)$ be a function of $n$ variables over ${\rm {\mathbb
F}}_{p} $, the quantum state $|\psi _{f} \rangle =p^{-\frac{n}{2} }
\sum _{x\in {\rm {\mathbb F}}_{p}^{n} } \zeta ^{f(x)} |x\rangle $ is
called logic state corresponding to $f(x)$, where $\zeta $ is a
primitive element in ${\rm {\mathbb F}}_{p} $. Specially, $|\psi
_{f} \rangle $ is called Boolean state corresponding to Boolean
function $f(x)$ if $p=2$.

 Denote quantum error as
$E_{\left(a,b\right)} =X\left(a\right)Z\left(b\right)$. Then,

\begin{equation} \label{GrindEQ__2_1_}
E(a,b)\left| {\psi _f } \right\rangle  = p^{ - \frac{n} {2}}
\sum\limits_{x \in \mathbb{F}_p^n } {\xi ^{f(x - a) + b(x - a)} }
\end{equation}

\noindent where $ \xi $ is a primitive element in ${\rm {\mathbb
F}}_{p}$, $ a = (a_1 , \cdots ,a_n ) \in \mathbb{F}_p^n $ and $ b =
(b_1 , \cdots ,b_n ) \in \mathbb{F}_p^n $, namely,
\begin{equation}
\left| {\psi _f } \right\rangle  \to E(a,b)\left| {\psi _f }
\right\rangle  \Leftrightarrow f(x) \to f(x - a) + b(x - a)
\end{equation}

Let ${\rm {\mathbb F}}_{p}^{n} $ be the vector space of dimension
$n$ over ${\rm {\mathbb F}}_{p} $ with the following inner product (
, ) defined by

\begin{equation} \label{GrindEQ__2_2_}
\left(a,b\right)=\sum _{i=1}^{n}a_{i} b_{i}
\end{equation}
for any $a=\left(a_{1} ,\cdots ,a_{n} \right)$, $b=\left(b_{1}
,\cdots ,b_{n} \right)$$\in {\rm {\mathbb F}}_{p}^{n} $. For
convenience, denote $\left(a,b\right)$ as $ a \cdot b $  .

For $K$ different vectors $\beta _{1} ,\cdots ,\beta _{K} $ and an
\textit{n}-variable function $f(x)$, $g_{i} (x)=f(x)+\beta _{i}
\cdot x$, $1\le i\le K$ are $K$ different functions. Further more,

\begin{equation} \label{GrindEQ__2_3_}
|\psi _{i} \rangle =p^{-\frac{n}{2} } \sum _{x\in {\rm {\mathbb
F}}_{p}^{n} } \zeta ^{g_{i} (x)} |x\rangle , 1\le i\le K
\end{equation}
are $K$ different logical states. Since,

\begin{equation} \label{GrindEQ__2_4_}
\sum _{x\in {\rm {\mathbb F}}_{p}^{n} } \zeta ^{f(x)-f(x)+(\beta
_{i} -\beta _{j} )\cdot x} =0,
\end{equation}
we have $\langle \psi _{i} |\psi _{j} \rangle =0$, namely, $|\psi
_{i} \rangle ,1\le i\le K$ are co-orthonogal.

\begin{definition}
The symmetrical distance between $a$ and $b$ is defined by

\begin{equation} \label{GrindEQ__2_5_}
W_{s} (a,b)=\# \{ i|1\le i\le n,(a_{i} ,b_{i} )\ne (0,0)\},
\end{equation}
where $a=\left(a_{1} ,\cdots ,a_{n} \right),b=\left(b_{1} ,\cdots
,b_{n} \right)\in {\rm {\mathbb F}}_{p}^{n} $.
\end{definition}

\begin{definition}\cite{proceeding15}
  Let $f(x)$ be an \textit{n}-variable Boolean function. The APC
distance of $f(x)$ is the minimum $W_{s} (a,b)$, where
$a=\left(a_{1} ,\cdots ,a_{n} \right),b=\left(b_{1} ,\cdots ,b_{n}
\right)\in {\rm {\mathbb F}}_{2}^{n} $ satisfying:

\begin{equation} \label{GrindEQ__2_6_}
\sum _{x\in {\rm {\mathbb F}}_{2}^{n} }
\left(-1\right)^{f(x)-f(x-a)-b\cdot x} \ne 0.
\end{equation}
\end{definition}

Xu \cite{proceeding16} generalized the definition of APC distance
for a Boolean function to logic function over ${\rm {\mathbb F}}_{p}
$ as following.

\begin{definition}\cite{proceeding16}
Let $f(x)$ be an \textit{n}-variable function over ${\rm {\mathbb
F}}_{p} $. The APC distance of $f(x)$ is defined by the minimum
$W_{s} (a,b)$, where $a=\left(a_{1} ,\cdots ,a_{n}
\right),b=\left(b_{1} ,\cdots ,b_{n} \right)\in F_{p}^{n} $
satisfying:

\begin{equation} \label{GrindEQ__2_7_}
\sum _{x\in {\rm {\mathbb F}}_{p}^{n} } \zeta ^{f(x-a)+b\cdot
x-f\left(x\right)} \ne 0,
\end{equation}
where $\zeta $ is a primitive element in ${\rm {\mathbb F}}_{p} $.
\end{definition}

\begin{definition}
The Hamming distance between $a$ and $b$ is defined by

\begin{equation} \label{GrindEQ__2_8_}
W_{H} (a,b)=\# \{ i|1\le i\le n,a_{i} \ne b_{i} \}
\end{equation}
with $a=\left(a_{1} ,\cdots ,a_{n} \right),b=\left(b_{1} ,\cdots
,b_{n} \right)\in {\rm {\mathbb F}}_{p}^{n} $.
\end{definition}

\section{Proof of Main Results}
In this section, let $f(x)$ be an \textit{n}-variable function with
APC distance $d'\ge 2$ over ${\rm {\mathbb F}}_{p} $ and  $ \beta _i
= \left( {\beta _{i 1} , \cdots ,\beta _{i n} } \right) \in {\Bbb
F}_p^n$ for all $1\le i\le K$.

For function $f(x)$ over ${\rm {\mathbb F}}_{p} $, constructing
quantum code $((n,K,d))_{p} $ by Lemma 1 is to find a group of
vectors, $\beta _{1} ,\cdots ,\beta _{K} $, with special
properties.The following theorem tells the properties of $\beta _{1}
,\cdots ,\beta _{K} $.


\begin{rom}
Quantum code $((n,K,d))_{p} $ spanned by \[\{ |\psi _{i} \rangle
=p^{-\frac{n}{2} } \sum _{x\in {\rm {\mathbb F}}_{p}^{n} } \zeta
^{f(x)+\beta _{i} x} |x\rangle |1\le i\le K\} \] is with following
properties:
\begin{enumerate}
\item  $d\le d'$,

\item  $\beta _{1} =\cdots =\beta _{K} =0\; $ for $d=d'$,

\item  $W_{H} \left(\beta _{i} ,\beta _{j} \right)\le k$ for all $d'=d-k$ if $0<k\le d'-2$.
\end{enumerate}
\end{rom}

\begin {proof}
We prove  $ d \le d'$ in two separate way firstly.

Case 1: $\exists 1\le i_{0} <j_{0} \le K$ satisfying $W_{H}
\left(\beta _{i_{0} } ,\beta _{j_{0} } \right)=t>0$. Then it is
reasonable to suppose $\beta _{2i} -\beta _{1i} \ne 0$ for all $1\le
i\le t$ and $\beta _{2i} =\beta _{1i} $ for all $t+1\le i\le n$.

If $t\ge d'$, set $u_0  = (1,\underbrace {0, \cdots ,0}_{n - 1}),v_0
= 0$. Thus,


\[W_{s} \left(u_{0} ,v_{0} -\beta _{i_{0} } +\beta _{j_{0} } \right)=t\ge d'.\]

\[d=\min \left\{W_{s} \left(u,v\right)|\exists 1\le i\le j\le K,W_{s} \left(u,v-\beta _{i} +\beta _{j} \right)\ge d'\right\}\]

\[\le W_{s} \left(u_{0} ,v_{0} \right)<d'.\]

If $t<d'$, set $u_0  = (\underbrace {0, \cdots ,0}_t,\underbrace {1,
\cdots ,1}_{d' - t},0, \cdots ,0)$, $v_0  = 0$. Then,

\[W_{s} \left(u_{0} ,v_{0} -\beta _{1} +\beta _{2} \right)=d'\]


\[d\le W_{s} \left(u_{0} ,v_{0} \right)=d'-t<d'\]

\noindent Therefore, \[d\le d'\] if $\exists 1\le i_{0} <j_{0} \le
K$ satisfying $W_{H} \left(\beta _{i_{0} } ,\beta _{j_{0} }
\right)=t>0$.

Case 2: $\beta _{i} =\beta _{j} $ for all $1\le i<j\le K$. Suppose
$W_{H} \left(\beta _{i} \right)=t$.

If $t\ge d'$, set $u_0  = (1,\underbrace {0, \cdots ,0}_{n - 1}),v_0
= 0$. Accordingly,

\[W_{s} \left(u_{0} ,v_{0} -\beta _{1} +\beta _{2} \right)=t\ge d',\]


\[d\le W_{s} \left(u_{0} ,v_{0} \right)<d'.\]

If $t<d'$, set $u_0  = (\underbrace {0, \cdots ,0}_t,\underbrace {1,
\cdots ,1}_{d' - t},0, \cdots ,0)$, $v_0  = 0$. As a result,

\[W_{s} \left(u_{0} ,v_{0} -\beta _{1} \right)=d',\]


\[d\le W_{s} \left(u_{0} ,v_{0} \right)=d'-t\le d'.\]

\noindent Therefore, \[d\le d'\] if $\beta _{i} =\beta _{j} $ for
all $1\le i<j\le K$.

We now prove  $ \beta _1  =  \cdots  = \beta _K  = 0 $ if
 $d= d'$.

First, we prove $\beta _{1} =\cdots =\beta _{K} $. Suppose $\exists
1\le i_{0} <j_{0} \le K$ satisfying $W_{H} \left(\beta _{i_{0} } ,
\beta _{j_{0} } \right)=t>0$. Hence, it is reasonable to suppose
$i_{0} =1,j_{0} =2$ and $\beta _{2i} -\beta _{1i} \ne 0$ for all
$1\le i\le t$, $\beta _{2i} -\beta _{1i} = 0$ for all $t+1\le i\le
n$.

 If $ t\ge d'$, set $u_0  = (1,\underbrace {0, \cdots
,0}_{n - 1}),v_0  = 0$. Consequently,

\[W_{s} \left(u_{0} ,v_{0} -\beta _{1} +\beta _{2} \right)=t>d',\]


\[d\le W_{s} \left(u_{0} ,v_{0} \right)=1<d'.\]

 If $t<d'$, set $u_0  = (\underbrace {0, \cdots
,0}_t,\underbrace {1, \cdots ,1}_{d' - t},0, \cdots ,0)$, $v_0  =
0$. Hence,

\[W_{s} \left(u_{0} ,v_{0} -\beta _{1} +\beta _{2} \right)=d',\]


\[d\le W_{s} \left(u_{0} ,v_{0} \right)=d'-t<d'.\]
\noindent A contradiction, therefore $W_{H} \left(\beta _{i} ,\beta
_{j} \right)=0$ for all $1\le i<j\le n$.

 Hence, $\beta _{1} =\cdots =\beta _{K} $. Denote $\beta
_{1} ,\cdots ,\beta _{K} $ as $\beta _{1} $.

Second, we prove $\beta _{1} =0$. Suppose $W_{H} \left(\beta _{1}
\right)=t>0$, thus, it is reasonable to suppose $\beta _{1i} \ne 0$
for all $1\le i\le t$ and $\beta _{2i} -\beta _{1i} = 0$ for all
$t+1\le i\le n$.

 If $t\ge d'$, set  $u_0  = (1,\underbrace {0, \cdots
,0}_{n - 1}),v_0  = 0$. As a result,

\[W_{s} \left(u_{0} ,v_{0} -\beta _{1} \right)=t,\]


\[d=\min \left\{W_{s} \left(u,v\right)|W_{s} \left(u,v-\beta _{1} \right)\ge
d'\right\}\] \[ \le W_{s} \left(u_{0} ,v_{0} \right)<d'.\]

If  $t<d'$, set $u_0  = (\underbrace {0, \cdots ,0}_t,\underbrace
{1, \cdots ,1}_{d' - t},0, \cdots ,0)$, $ v_0  = 0$. Consequently,

\[W_{s} \left(u_{0} ,v_{0} -\beta _{1} \right)=d',\]


\[d\le W_{s} \left(u_{0} ,v_{0} \right)=d'-t<d'.\]
\noindent A contradiction, therefore, $W_{H} \left(\beta _{1}
\right)=0$.

 This completes the proof of property $2)$.

We now prove property $3)$. Suppose $\exists 1\le i_{0} <j_{0} \le
K$ satisfying $W_{H} \left(\beta _{i_{0} } ,\beta _{j_{0} }
\right)\ge k+1$. Then it is reasonable to suppose $i_{0} =1,j_{0}
=2$. Denote $W_{H} \left(\beta _{1} ,\beta _{2} \right)=t$, where
$t\ge k+1$. Thus it is reasonable to suppose $\beta _{1i} \ne \beta
_{2i} $ for all $1\le i\le t$ and $\beta _{2i} -\beta _{1i} = 0$ for
all $t+1\le i\le n$.

 If $t\ge d'$, set $u_0  = (1,\underbrace {0, \cdots ,0}_{n
- 1}),v_0  = 0$. Hence,

\[
W_s \left( {u_0 ,v_0  - \beta _1  + \beta _2 } \right) = t \ge d'.
\]
\[
 d\le W_s (u_0 ,v_0 ) < d' - k.
\]

If $t<d'$, set $ u_0  = (\underbrace {0, \cdots ,0}_t,\underbrace
{1, \cdots ,1}_{d' - t},0, \cdots ,0),v_0  = 0 $. Accordingly,

\[W_{s} \left(u_{0} ,v_{0} -\beta _{1} +\beta _{2} \right)=t\ge d',\]


\[d\le W_{s} \left(u_{0} ,v_{0} \right)=d'-t\le d'-k-1.\]

\noindent A contradiction, therefore $W_{H} \left(\beta _{i} ,\beta
_{j} \right)\le k$ for all $1\le i<j\le K$ if $0<k\le d'-2$.

This completes the proof of  Theorem $1$.
\end {proof}

\begin{remark}

It can be easily seem from Theorem $1$  that if the following
conditions satisfy:

\begin{enumerate}
\item  There exists an \textit{n}-variable function with APC distance $d'\ge 2$ over ${\rm {\mathbb F}}_{p}
$,

\item  A group of vectors $\beta _{1} ,\cdots ,\beta _{K} $ over ${\rm {\mathbb F}}_{p}^{n} $ satisfy $W_{H} \left(\beta _{i} ,\beta _{j} \right)\le k$ for all $1\le i<j\le K$.
\end{enumerate}

\noindent  Quantum code $((n,K,d'-k))_{p} $ can be constructed by
Lemma $1$.
 \end{remark}

In the following theorem, we are going to deal with the parameter
$K$.

\begin{rom}
If quantum code $((n,K,d))_{p} $ is spanned by $\{ |\psi _{i}
\rangle =p^{-\frac{n}{2} } \sum _{x\in {\rm {\mathbb F}}_{p}^{n} }
\zeta ^{f(x)+\beta _{i} x} |x\rangle |1\le i\le K\} $.  Then

\[K=\left\{\begin{array}{l} {1\; ,\; \; \; \; \; \; d=d'} \\ {\le p,\; \; \; d=d'-1} \\ {\le \max p^{k-2} (1+n(p-1),p^{2} )\; \; ,\; d=d'-k} \end{array}\right. ,\]
where $2\le k\le d'-2$.
\end{rom}

\begin {proof}
\begin{enumerate}
\item For $ d=d' $, it can be deduced from Theorem 1 that
\[\beta _{1} =\cdots =\beta _{K} =0.\]
Thus,

\[\textit{K}=1.\]

\item  For $d=d'-1$, let $W_{ij} =W_{H} (\beta _{i} ,\beta _{j}
)$ for all $1\le i<j\le n$.

Suppose $K>p$. Then there exists $1\le i_{0} <j_{0} \le K$
satisfying $W_{i_{0} j_{0} } \ge 2$, a contradiction, thus

\[K\le p.\]

\item  Denote $C_{n}^{t} $ as the number of vectors where the
Hamming distance between each other is no more than $t$.

For $k=2$, since $W_{H} (\beta _{i} ,\beta _{j} )\le 2$ for all
$1\le i<j\le K$  by  Theorem $2$.

Case 1: If $\beta _{1} ,\cdots ,\beta _{K} $ are the same in $n-2$
bits. It can be deduced that $\beta _{1} ,\cdots ,\beta _{K} $ are
different in at most 2 bits, hence,

\[K\le p^{2} .\]

Case 2: If that $\beta _{1} ,\cdots ,\beta _{K} $ are the same in
$n-2$ bits doesn't satisfy, then, \textit{K} is the maximal when the
different bits are all \textit{n} bits. Thus,

\[K\le (p-1)n+1\]

\noindent Therefore, $K\le \max \{ p^{2} ,(p-1)n+1\} $ for $d=d'-2$.

For $3\le k\le d'-2$, since $W_{H} (\beta _{i} ,\beta _{j} )\le k$
by Theorem $1$ for all $1\le i<j\le K$. Thus,

\[K=C_{n}^{k} \le pC_{n-1}^{k-1} \le \cdots \le p^{k-2} C_{n-k+2}^{2} \]

\[\le \max p^{k-2} \{ 1+(n-k+2)(p-1),p^{2} \} \]
This completes the proof of  Theorem $2$.
\end{enumerate}
\end {proof}

\begin{remark}
It can be inferred from Theorem $1$ and Theorem $2$ that for an
\textit{n}-variable function with APC distance $d'\ge 2$ over ${\rm
{\mathbb F}}_{p} $, quantum code with parameters $((n,K,d))_{p} $
can be constructed by Lemma $1$ where $d\le d'$. Furthermore, if
$d=d'-k,0\le k\le d'-2$, then $\beta _{1} ,\cdots ,\beta _{K} $
should satisfy $W_{H} (\beta _{i} ,\beta _{j} )\le t$ for all $1\le
i<j\le K$. At the same time, we obtain the maximal $K$.
\end{remark}

\section{Basic States and Equivalent Conditions of Constructing Quantum MDS Codes}

\subsection{The basic states of the constructed quantum code}

In this subsection, denote  $\beta _i$ as $ \beta _i  = \left(
{\beta _{i1} , \cdots ,\beta _{in} } \right) $.

 For an \textit{n}-variable function with APC distance $d'$ over ${\rm
{\mathbb F}}_{p} $ and $\beta _{1} ,\cdots ,\beta _{K} $, quantum
code $((n,K,d))_{p} $ can be constructed by Lemma $1$. The basic
states of the constructed quantum code can be stated as following:

If $p\ge n-k+1$, then \[p^{k} \ge p^{k-2} +p^{k-2} (p-1)(n-k+2).\]
Let \[K=p^{k} .\]  At this time, we set  $\beta _{1} ,\cdots ,\beta
_{K} $ be vectors that the first $k$ bits run all over ${\rm
{\mathbb F}}_{p}^{k} $ and the last $n-k$ bits are zeros.
 Namely,
 \begin{equation}
 \beta _{ij}  \in \mathbb{F}_p  ~for~ 1\le j \le k
  \end{equation}
  \begin{equation}
   \beta_{ij}=0 ~for~ k+1\le j \le n
 \end{equation}
where $1 \le i \le p^{k}$. It can be checked that $W_{H} (\beta _{i}
,\beta _{j} )\le k$ for all $1\le i<j\le p^{k} $, thus, the space
spanned by formula $(4)$ corresponding to $\beta _{1} ,\cdots ,\beta
_{K} $ satisfying formula (10) and  (11) is a quantum code with
parameters ${\rm ((}n,K,d'-k{\rm ))}_{p} $.

If $p<n-k+1$,  then $p^{k-2} +p^{k-2}
\left(n-k+2\right)(p-1)+1>p^{k} $. Let \[K=p^{k-2} +p^{k-2}
\left(n-k+2\right)(p-1).\]  At this time, we set $\beta _{1} ,\cdots
,\beta _{K} $ be vectors that the first $k-2$ bits run all over
${\rm {\mathbb F}}_{p}^{k-2} $ , the $k+l-2$ -th bit run all over
${\rm {\mathbb F}}_{p} \backslash \left\{0\right\}$,  $ 1 \le l \le
n - k + 2$. Namely,
\begin{equation}
 \beta _{ij} \in \mathbb{F}_p ~ for ~1\le j \le k-2
\end{equation}
\begin{equation}
 \beta _{i~ k + l - 2} \in \mathbb{F}_p \backslash \{ 0\}  ~for~ 1\le
l \le n-k+2
\end{equation}
and the rest bits are all zeros. It can be easily checked that
\[W_{H} (\beta _{i} ,\beta _{j} )\le k-2+2=k\] for all $1\le i<j\le
K$, thus, the space spanned by formula $(4)$ corresponding to $\beta
_{1} ,\cdots ,\beta _{K} $ satisfying formula (12) and formula (13)
is a quantum code with parameters \[((n,p^{k-2} +p^{k-2}
(p-1)(n-k+2),d'-k))_{p} .\]

\subsection{The equivalent conditions of constructing quantum MDS codes}
Theory of quantum code has quantum singleton bound as classical
code. Quantum codes saturating quantum Singleton Bound are quantum
MDS codes. The following theorem presents the equivalent conditions
of quantum MDS codes constructed by Lemma 1.


\begin{theorem}
 Quantum code ${\rm ((}n,K,d'-k{\rm ))}_{p} $ is
constructed by Lemma 1, where $d'-k\le \frac{n}{2} +1$. Then it
saturates quantum Singleton Bound if and only if the following
conditions satisfy:

\begin{enumerate}
\item  If $k=0$, then there exists an \textit{n}-variable function
over ${\rm {\mathbb F}}_{p} $ with APC distance  $d'$ over ${\rm
{\mathbb F}}_{p} $, where $d'=\frac{n}{2} +1$ and $n$ is even,

 \item  If $k=1$, then there exists an \textit{n}-variable
function with APC distance $d'$ over ${\rm {\mathbb F}}_{p} $, where
$d'=\frac{n}{2} +1$,

 \item  If $2\le k\le d'$ and $p\ge n-k+1$, then there
exists an \textit{n}-variable function with APC distance $d'$ over
${\rm {\mathbb F}}_{p} $, where $2d'=n+k+2$,

 \item  If $2\le k\le d'$and$p<n-k+1$, then there exists an
\textit{n}-variable function with APC distance $d'$ over ${\rm
{\mathbb F}}_{p} $, where $p^{k-2} +p^{k-2}
\left(n-k+2\right)(p-1)=p^{n-2(d'-k)+2} $.
\end{enumerate}
\end{theorem}

\begin{proof}
 Let quantum code $((n,K,d'-k))_{p} $ be constructed by
Lemma $1$.

\begin{enumerate}
\item  If $k=0$, then

\[K=1\]

by Theorem 2. Thus, the quantum code saturates Quantum Singleton
Bound if and only if

\[n-2d'+2=0.\]

\item  If $k=1$, we get

\[K\le n(p-1)+1\]

by Theorem 2. Thus, the quantum code saturates Quantum Singleton
Bound if and only if

\[n(p-1)+1=p^{n-2d'+4} .\]

\item  If $2\le k\le d'$ and $p\ge n-k+1$,

\[K\le p^{k} \]

by Theorem 2. Thus, the quantum code saturates Quantum Singleton
Bound if and only if

\[k=n-2\left(d'-k\right)+2\Leftrightarrow 2d'=n+k+2.\]

\item  If $2\le k\le d'$ and $p<n-k+1$,

\[K<p^{k-2} +p^{k-2} \left(n-k+2\right)(p-1)\]

by Theorem 2. Thus, the quantum code saturates Quantum Singleton
Bound if and only if

\[p^{k-2} +p^{k-2} \left(n-k+2\right)(p-1)=p^{n-2(d'-k)+2}.\]
\end{enumerate}

 This completes the proof of this Theorem .
\end{proof}

\section{Conclusion}
Ref. \cite{proceeding16} presented a new way to construct quantum
error correcting  codes. Quantum error correcting codes can be
constructed by use of logic functions with \textit{n} variables and
APC distance $d'\ge 2$ over ${\rm {\mathbb F}}_{p} $.
 The minimum distance of the constructed quantum code is
 $d=d'-t(0\le t\le d'-2)$. We can also get the maximal dimension of
 the corresponding space. In this paper, we also give the basic
 states and the equivalent conditions for existence of quantum MDS codes.

It can be seem that logic functions with favorable APC distance play
a key role in logic construction for quantum codes. The presented
paper is to re-cast the construction of QECCs as a problem of
construction logic function with favorable APC distance. Ref [17]
proposed a quadratic residue construction for Boolean function with
favorable APC distance. For an \textit{n}-variable function over
${\rm {\mathbb F}}_{p} $, how to compute the APC distance fast is
still a problem to be researched.

\section*{Acknowledgment}
This work is supported by the NFS of China under Grant number
60403004 and the Outstanding Youth Foundation of Henan Province
under Grant No.0612000500.


\begin{thebibliography}{1}

\bibitem{proceeding1} P. W. Shor, `` Scheme for Reducing Decoherence in Quantum Computer Memory,'' \emph{Phys. Rev.\ A.}  54 (2),  pp.~ 1098--1105, 1995.

\bibitem{proceeding2}C. H. Bennettt, D. P. DiVincenco, J. A. Smolin  and
W. K. Wootters, `` Mixed state entanglement and quantum error
correction,"  \emph{Phys.\ Rev.} 54 (5), pp.~ 3824--3851, 1996.

\bibitem{proceeding3}E. Knill and R. Laflamme, `` A Theory of quantum error-correcting
 code saturating quantum Hamming Bound," \emph{Phys. Rev. A.}  55, pp.~ 900--911, 1997.

\bibitem{proceeding4}A. M. Steane, `` Simple quantum error correcting
codes,"
 \emph{Phys. Rev. Lett.}  77, pp.~793--797, 1996.

\bibitem{proceeding5}D. Gottesman, `` Theory of fault-tolerant quantum
computation," \emph{ Phys.\ Rev.\ A.}
 57, pp.~ 127--137, 1998.

\bibitem{proceeding6}C. H. Bennett  and G. Brassard, `` Quantum
cryptography: public key distribution and coin tossing,''
\emph{Proceedings of IEEE International Conference on Computers,
Systems, and Sig-nal Processing,}
 pp.~175--179,  1984.

\bibitem{proceeding7}S. Glancy,  E. Knill  and H. M. Vasconcelos,``
Entanglement purification of any stabilizer state,''
 \emph{Phys.\ Rev.\ A.}  74, no. 032319, 2006.

\bibitem{proceeding8} A. Ambainis  and  D. Gottesman, ``The minimum distance problem
for two-way entanglement purification,''
 \emph{IEEE Trans.\ Inform.\ Theory.} 52,  pp.~748--753, 2006.

\bibitem{proceeding9}A. R. Calderbank, E. M. Rains,  P. W.  Shor and N. J. A.
Sloane, `` Quantum error correction via codes over  $ \mathbb{F}_4
$,'' \emph{IEEE Trans.\ Inform Theory} 44,  pp. ~1369--1387,  1998.

\bibitem{proceeding10} D. Schlingemann  and  R. F. Werner,``Quantum error correcting codes associated with graphs,''  \emph{Phys.\ Rev.\ A.} 65, 012308, 2002.

\bibitem{proceeding11}E. M. Rain, `` Nonbinary quantum code,'' \emph{ IEEE Trans.\ Inform Theory} 45,  pp.~1827--1832, 1999.

\bibitem{proceeding12} K. Q. Feng, `` Quantum codes $ [[6,2,3]]_p $ and $ [[7,3,3]]_p $ ($p\ge 3$)  exist,'' \emph{IEEE Trans.\ Inform Theory} 48 (8),
pp.~2384--2391, 2002.

\bibitem{proceeding13} T. L. Liu,  `` On  construction for nonbinary cyclic quantum code via graph,'' \emph{China Science\ Inform Theory.\ E.} 35 (6), pp.~588--596, 2005.

\bibitem{proceeding14} V. Aggarwal  and R. Calderbank, ``
Boolean functions, projection operators and quantum error correction
codes,'' \emph{IEEE Trans.\ Inform Theory.}, 54 (4) PP.~1700--1707,
2008.

\bibitem{proceeding15} L. E.  Danielsen, ``  On self-dual quantum codes, graphs, and Boolean
functions,''
 http://arxiv.org/abs/quant-ph/0503236, 2005.12.

\bibitem{proceeding16}Y. J. Xu,  ``Logic function and quantum code,''
http://arxiv.org/abs/quant-ph/0712.3605v4, 2008.01.

\bibitem{proceeding17} L. E. Danielsen, `` Aperiodic Propagation Criteria for Boolean Functions,''
  \emph{In Information and Computation} 204 (5), pp.~741--770, 2006.


\end{thebibliography}
\end{document}